\documentclass[]{aa}

\usepackage{graphicx,epsf,psfig,epsfig,times,amsmath,amssymb}

\usepackage{natbib}
\bibpunct{(}{)}{;}{a}{}{,}

\def\comptt{{\sc Comptt}}
\def\comptb{{\sc Comptb}}

\def\cm2{cm$^{-2}$}
\def\s1{s$^{-1}$}

\def\sax{{\it BeppoSAX}}
\def\xte{{\it RXTE}}
\def\integral{{\it INTEGRAL}}
\def\isgri{{\it ISGRI}}
\def\xspec{{\rm  XSPEC}}

\def\ktbb{kT_{\rm bb}}

\def\kts{kT_{\rm s}}
\def\kte{kT_{\rm e}}

\def\qtot{Q_{\rm tot}}

\def\qd{Q_{\rm disk}}
\def\qcor{Q_{\rm cor}}

\begin{document}

\title{On the stability of the thermal Comptonization index in neutron star low-mass X-ray binaries in their different 
spectral states}

\author{R. Farinelli\inst{1} \& L. Titarchuk\inst{1, 2}}
\offprints{R. Farinelli, farinelli@fe.infn.it}
\institute{
Dipartimento di Fisica, Universit\`{a} di Ferrara, Via Saragat 1, 44100 Ferrara, Italy
\and NASA/GSFC, Greenbelt, MD 20771}

\abstract 
{Most of the spectra of neutron star low mass X-ray binaries (NS LMXBs), being them persistent or transient, 
are characterized by the presence of a strong thermal Comptonization bump, thought to originate in the 
transition layer (TL) between the accretion disk and the NS surface.  The observable quantities which characterize this component dominating  the emission below 30 keV, are the spectral index $\alpha$ and the rollover energy, both related to the electron temperature and optical depth of the plasma.}
{Starting from observational results on a sample of NS LMXBs in different spectral states, we formulate the problem of  X-ray spectral formation in the TL 
of these sources. We predict a stability of  the thermal Comptonization spectral index in different spectral states if 
the energy release in the  TL is much higher than the intercepted flux coming from  the accretion disk. }
{We use an equation for the energy balance and the radiative transfer diffusion equation for a slab geometry in the 
TL, to derive  a formula  for the thermal Comptonization index $\alpha$.  
We show that in this approximation  the TL electron temperature $\kte$ and optical depth $\tau_0$ can be written as a function of the  energy flux from the disk intercepted by the corona (TL) and  that in the corona itself, $\qd/\qcor$. As  spectral index $\alpha$ depends on $\kte$ and $\tau_0$, this in turn leads to a relation $\alpha=f(\qd/\qcor)$, with $\alpha \sim 1$ when   $\qd/\qcor \ll 1$.
 }
{We show that the observed spectral index $\alpha$ for the sample of sources here considered lies in a belt around $1 \pm 0.2$ a part
for the case of GX 354--0. Comparing our theoretical predictions with  observations, we claim that this result, which is 
consistent with the condition $\qd/\qcor \ll 1$, can  give us  constraints on the accretion geometry of these systems, an issue that seems difficult to be solved using only the spectral analysis method.}{}

\authorrunning{Farinelli  \& Titarchuk }
\titlerunning{Thermal Comptonization index in NS LMXBs}
\maketitle

\section{Introduction}
\label{introduction}

It is well known since the 80's that the spectra of low-mass \mbox{X-ray} binaries
(LMXBs) hosting a neutron star (NS) can be described up to about 30 keV by
 a two-component model representing the contribution  from different
emitting regions of the system.
However, the interpretation of the spectra is not unique 
as demonstrated using  the variety of different models used across the years.

Before the \sax\ and \xte\ era,  two concurring models were classically used to described X-ray emission in LMXBs. In the so-called ``eastern model'' \citep{mitsuda84, mitsuda89} the spectra were fitted by the sum of a soft
blackbody (BB) emission (actually modeled by a multi-colour disk BB spectrum) 
attributed to the accretion disk, plus a hotter simple or Comptonized BB
claimed to originate close to the NS surface.
On the other hand, in the ``western model'' interpretation \citep{white86, white88}, the direct BB component was attributed to the NS surface, while 
an unsaturated Comptonization spectrum was thought to originate from a hot corona above  the inner accretion disk, which supplies most of the soft seed photons for Comptonization.

In fact, even after the advent of \sax\ and \xte,  the persistent emission of NS LMXBs was described by the sum of a BB component plus a thermal Comptonization (TC) spectrum, usually described by  the \xspec\  \comptt\ model \citep{t94, ht95}.
Despite the significant improvement in our knowledge of the source spectral properties by means of the broad-band observations, the BB+TC model resulted and subjected to a dichotomy. Indeed, both of  the  cases where the temperature of the direct BB spectrum $\ktbb$ is lower \citep[e.g.,][]{ds00a, ds00b, ds01, oosterbroek01, gd02, lavagetto04, paizis05} or higher \citep[e.g.,][hereafter F08]{paizis05, farinelli07, farinelli08} than that of the thermally Comptonized seed photons, $\kts$,  provide in general equally acceptable good fits. 

The consequences of these results were the interpretation of the BB emission
as due either to the accretion disk ($\ktbb < \kts$) or to the NS surface ($\ktbb > \kts$).
Moreover, in addition to the persistent X-ray emission,  \sax\ \citep[e.g.,][]{ds00b,ds02} \xte\  \citep{damico01, ds06} and later also \integral\ \citep[][hereafter P06]{paizis06} allowed the possibility to discover in bright NS LMXBs a transient powerlaw (PL)
X-ray emission above 30 keV .

Motivated by the need to put some order and give an unified scenario in the different NS systems spectral states,  Paizis et al. (2006, hereafter P06) performed a systematic observational campaign with the \isgri\ (20-200 keV) monitor onborad \integral. Using long-term average spectra and including former \integral\ results on GX 354--0 \citep{falanga06}, P06 classified NS LMXBs into four main states: the \emph{hard/PL}, \emph{low/hard}, \emph{intermediate} and \emph{soft}.
The high-energy ($>$ 20 keV) spectra in different sources were interpreted
by P06 as the result of the interplay between thermal and bulk Comptonization processes whose relative efficiency is ultimately dictated by the mass accretion rate.
At high energies (where the direct BB component is negligible), the \emph{hard/PL} state spectra can be fitted with a simple PL component; the \emph{low/hard} spectra by a TC spectrum of soft ($\la 1$ keV) BB-like photons
off an electron population with $\kte \sim 20-30$ keV and $\tau_0 \la 3$; the \emph{intermediate} state spectra by TC spectrum with $\kte \sim 3-5$ keV and $\tau_0\ga 5$ plus a PL component with photon index $\Gamma \sim 2-3$; the  \emph{soft} state spectra by a TC component similar to the \emph{intermediate} state but without the high-energy  X-ray tail.

From the observational point of view, the quantities directly measurable  in the data 
are the cut-off energy $E_c$ of the dominating TC bump and the spectral slope, parametrized through the energy index $\alpha$ (=$\Gamma -1$). 
For pure TC spectra, $\alpha$ is tightly correlated with the plasma temperature $\kte$ and optical depth $\tau_0$ \citep[][hereafter TL95]{st80, tl95}, while in the
presence of a converging flow (bulk motion), the shape of the velocity field also determines the emerging spectral slope \citep[][F08]{tmk97, LT99}.
In mathematical terms, $\alpha$ represents the index of the system Green's function (GF), namely as it responds to monochromatic line injection. The resulting  emerging spectrum is then obtained as a convolution of the GF with the input seed photon spectrum.
We also emphasize that the \emph{hard/PL}, \emph{hard} and \emph{soft} states spectra of NS LMXBs are characterized by the presence of just \emph{one} Comptonization index (see Fig. 4 in P06). In the first case, its value is interpreted as a result of a mixed TC plus BC process, while in the latter two cases, $\alpha$ is derived from pure TC. On the other hand, the \emph{intermediate} state spectra show \emph{two} Comptonization indexes, 
one is  related to the persistent TC component and the other one characterizes the transient PL-like hard X-ray  emission.

In Section \ref{sect_alpha} we give an overview of the theoretical and observational issues related to  
spectral evolution in X-ray binary systems hosting a NS or a black hole (BH) outlining the differences among the two
classes of sources. We subsequently report on results related to the observed TC index $\alpha$ for a sample of  NS sources.
In Section \ref{model} we propose a theoretical model based on diffusion formalism for radiative transfer in order
to explain the observational results. In Section \ref{results} we discuss the comparison between theory and
data, while in Section \ref{conclusions} we draw our conclusions and give future observational perspects.

\section{Spectral index evolution in NS systems}
\label{sect_alpha}

Starting from the considerations of the previous section, it is important to make a comparison between the index 
evolution in accreting BH and NS sources.
The spectral state of BH sources may be generally divided into \emph{low/hard} state (LHS), where the spectrum is dominated by a TC component with electron temperature $\kte \sim$ 60-100 keV, \emph{intermediate}  state (IS), with a BB bump (presumably coming from the accretion disk) and a superposed PL high-energy component, and \emph{high/soft} state (HSS) where the BB component gets even stronger and the PL emission gets steeper.

Actually, in BH systems generally \emph{one high energy photon index} $\Gamma$ is observed
[see  e.g.  recent results on GRS 1915+015 by Titarchuk \& Seifina (2009) and \cite{st09}] and
its value evolves  with the source mass accretion rate.
The latter, which is not a direct observable quantity, may be inferred in an indirect way by means of the 
the normalization of the disk flux $N_{\rm disk}$.
In  a $\Gamma$ vs $N_{\rm disk}$/Quasi Periodic Oscillation (QPO) frequency diagram, it was found \citep{st09, ts09,mtf09}
that the photon index $\Gamma$ progressively increases from $\sim$ 1.6-1.8 as the source moves from the \emph{low/hard} to the \emph{high/soft} states, until it reaches saturation around
$\Gamma \sim 2.2-2.4$ (depending on the source). 
This interpretation of the observed BH spectra have been performed using  ``so called''  the BMC model \citep{tmk97}, whose emerging spectral shape is described by the formula
\begin{equation}
F(E)=\frac{C_N}{A+1} [BB(E)+ A \times G(E,E_0) \ast BB(E_0)].
\label{bmc}
\end{equation}
In formula (\ref{bmc}), the first term of the right-hand side represents
the direct seed photon BB-like spectrum, while the second one gives its modification
due to  Comptonization (convolution with GF).
The hardness of the spectrum is dictated by the energy index $\alpha$ of the
GF which, in BMC, is a broken-PL with $G(E,E_0) \propto E^{\alpha+3}$ for
$E < E_0$ and $G(E,E_0) \propto E^{-\alpha}$ for $E > E_0$,  where $E_0$ is the monochromatic
input energy. It is important to keep in mind that the GF in BMC  does not contain the exponential spectral rollover,
so when the latter is observed in the X-ray spectrum of a source, it can be taken into account by multiplying 
BMC with an exponential e-folding factor $\propto e^{-E/E_c}$.

The index saturation observed in BH sources as they move to the HSS finds a natural explanation in the framework of a bulk-dominated scenario 
\citep[see mathematical proof of this statement in e.g.][]{ts09}.
Moreover, it is interesting to note  that when the sources are in the 
 LHS, the points in the $\Gamma$ vs $N_{\rm disk}$/QPO diagram form a plateau around $\Gamma \sim$1.5 before the rising phase.
This clear mapping of the energy (or photon) index evolution is possible in BH sources as they are generally strongly variable 
and their X-ray spectrum allows a one-to-one correspondence between the spectral state and the energy 
index $\alpha$ (or $\Gamma$)  using spectral modeling according to  formula (\ref{bmc}). One of the advantages of using
BMC model is that it is a \emph{generic} Comptonization model, namely it allows to map the spectral
evolution of sources through the Comptonization index $\alpha$, which is a direct measurable quantity, no matter
which are the underlying physical conditions. The theoretical interpretation of the source spectral formation
is in fact related to a later scientific discussion.

For NS sources however, situation is less straightforward to face.
In fact,  most variable sources such as, e.g.,  \mbox{GX 354--0} \citep{falanga06}
or 4U 1608--52 \citep{gd02} when moving from  the \emph{hard} state to the \emph{soft}
state do exhibit pure TC spectra, with electron temperature $\kte$ progressively decreasing and optical depth $\tau_0$ increasing, respectively.
On the other hand, bright LMXBs of the GX class  such as the classical six known
Z sources (Sco X--1, GX 17+2, Cyg X--2, GX 340+0, GX 5--1 and GX 349+2) and, 
more recently, GX 13+1, show only a small evolution of their persistent X-ray continuum
(dominated by the strong TC bump with $\kte \sim$ 3-5 keV and $\tau_0 \gg 1$). In addition they  are characterized by the presence of a transient hard X-ray PL-like component
(\emph{intermediate} state).

Other persistently bright sources such as GX 3+1, GX 9+1 and GX 9+1
have been only  observed  in the \emph{soft} state with no evidence of hard X-ray tails.
In fact, mapping the evolution of the transient hard X-ray tail of NS sources in the
\emph{intermediate} state has not been yet possible because of the insufficient statistics
available at high energies. Some attempts \citep{ds00a,ds02,ds06} were actually undertaken  to establish changes in the  intensity  of the transient hard tail by  fitting it  with a simple PL and, when statistics was poor, fixing the index 
allowing to
vary only its flux. But nothing could be concluded about the index evolution.
Thus, at the present status of knowledge, serious investigations can be performed
only on the evolution of the spectral index related to the persistent TC component.
This was done for the first time by \citet[][hereafter TS05]{ts05}, who performed
a systematic analysis of the variable NS X-ray binary 4U 1728-34 from the \emph{hard} state
to the \emph{soft} state using observations from the Proportional Counter Array (PCA, 3-30 keV) onboard \xte.
The source spectra were fitted with a two-BMC model, with $A \gg 1$ 
(see formula  [\ref{bmc}]) and with the GF spectral index fixed equal for both BMC components. Thus, the model used by TS05 actually was
$$F(E)= C_{ns}G(E,E_0) \ast BB(E_0) +  C_{disk}G(E,E_0) \ast BB(E_0).$$

The main result found by TS05 was that as the source moved from
the \emph{hard} to the \emph{soft} state the index $\Gamma$ ($=\alpha+1$) progressively increased with no evidence of saturation, unlike the BH case. Actually, the final \emph{soft} state of  4U 1728-34
was represented by the sum of two BB component,  because  for $\alpha \gg 1$, $G(E,E_0) \ast BB(E_0) \approx BB(E).$

This result however needs a revision. Namely,  fitting the NS LMXBs \emph{soft} state
 spectrum with a two-BB model can be possible because of the lack of data below 3 keV, which is of key importance. 
The  limited broad-band  resolution allows however the possibility to have 
different models with same good fitting results. For example,  recent results of the analysis  of PCA/\xte\ data for 4U 1608--52 by Ding et al. (2010 in preparation) show that the \emph{soft} state spectrum of the source
can be fitted either by a two-BB model or by a single TC model (e.g., \comptb)
with $\alpha \sim$ 1. Note also that Falanga et al. (2006) fitted the \emph{soft state}
spectrum of GX 354--0 with a DBB+\comptt\ model, from which the inferred $\alpha$-value
is again $\sim$ 1.
Actually, when looking at the \sax\ results obtained over years of observations of NS LMXB sources, it results that the \emph{soft} state spectra of these systems, 
rather than 2 BBs, need to be described by the sum of a BB component
plus an unsaturated TC spectrum with cut-off energy below 10 keV.

\subsection{Observational results}

 We considered a sample of sources taken from the literature and for which we can make a fiducial measurement of $\alpha$.
In our choice of the sources, we adopted the criterion to consider those in which
$\alpha$ was determined either directly from the fit, as it can be done using the 
\comptb\ model (where $\alpha$ is a free parameter, see F08), or can be derived
from the temperature and optical depth obtained by the \comptt\ model \citep{t94}.

The sources belonging to the first case are \mbox{Sco X--1}, \mbox{GX 17+2}, \mbox{Cyg X--2}, \mbox{GX 340+0}, \mbox{GX 3+1}
(see Table 2 in F08) and \mbox{GS 1826--238} \citep[see Table 2 in][]{cocchi10}.
For \mbox{GX 349+2} \citep{ds00a} and \mbox{GX 354--0} \citep{ds00b} we derived the value of spectral index  
$\alpha$ using equation for the non-relativistic regime 
(see Eq. [22] in Titarchuk \& Lyubarskij 1995, hereafter TL95)
\begin{equation}
\alpha=-\frac{3}{2}+\sqrt{\frac{9}{4}+\frac{\beta}{\Theta}}, 
\label{alpha_general}
\end{equation}
where $\Theta \equiv \kte/m_e c^2$ and  $\beta$-parameter
defined in formula (17) of TL95 for spherical geometry
as it was assumed by the authors. In the case of X1658--298, \cite{oosterbroek01} assumed a slab geometry, thus $\alpha$ was obtained from equation (\ref{alpha_general}) and formula (17) of TL95 for a slab geometry. 
For 1E 1724--3045, \cite{barret00} report the best-fit value of optical depth $\tau_0$ of Comptonization region both for the case of spherical and slab geometry. We checked that the two derived values of $\alpha$ are perfectly consistent.
The errors on $\alpha$ for sources for which $\kte$ and $\tau_0$ were reported, have been computed considering that the 
function  $\alpha[\kte, \beta(\tau_0)]$ gets its absolute minimum and maximum 
values at the boundary of the box of its domain delimited by the minimum and  maximum value ($kT^{\rm min}_{\rm e}$, $kT^{\rm max}_{\rm e}$) and ($\tau^{\rm min}_0$, $\tau^{\rm max}_0$)
obtained in computing the errors at 90\% confidence level for the electron temperature and optical depth by \xspec.

\begin{figure}
\centering
\includegraphics[width=6cm, angle=-90]{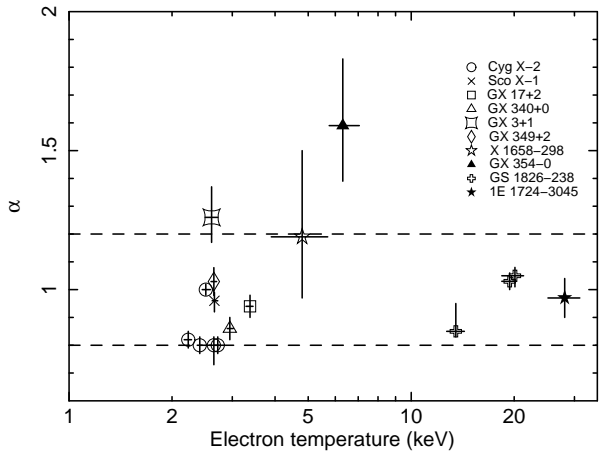}
\caption{Thermal Comptonization index $\alpha$ for sources in different spectral states as a 
function of the electron temperature $\kte$. 
Reference papers: \mbox{Cyg X--2}, Farinelli et al. (2009); \mbox{Sco X--1}, \mbox{GX 17+2}, 
\mbox{GX 340+0}  and \mbox{GX 3+1}, Farinelli et al. (2008); \mbox{GX 354--0}, Di Salvo et al. (2000b); 
\mbox{GX 349+2}, Di Salvo et al. (2001), \mbox{X 1658--298}, Oosterbroek et al. (2001); 
\mbox{GS 1826--238}, Cocchi et al. (2010); \mbox{1E 1724--3045}, Barret et al. (2000).} 
\label{alpha_data}
\end{figure}

In Figure \ref{alpha_data} we report the measured values of $\alpha$ for this 
sample of sources as a function of the electron temperature $\kte$.
This parameter can be considered indeed a good tracer of the source spectral state because 
$\kte$ decreases when sources move from the \emph{hard} to \emph{soft} state as a result of a more efficient electron cooling by the enhanced seed photon supply.
Moreover, the electron  temperature $\kte$   is a directly measurable quantity because it is related to  the cut-off energy of the spectrum and it presents the advantage of being distance-independent.
On the other hand, the instrumental energy-band coverage and
accumulation time  can  play some role in biasing the measured index value.
For instance, \cite{falanga06} performed a systematic analysis of GX 354--0 as a function
of its position on the hardness-intensity diagram when the source moved from the \emph{hard} to \emph{soft} state. 
The authors fitted the 3-100 keV spectrum with a multicolour disk BB \citep[DBB,][]{mitsuda84} plus \comptt\ model using the slab geometry.
We computed the derived $\alpha$-values from their best-fit parameters but the trend was not monotonic, covering the range $\sim$ 1-3 and 
 reflecting such a behaviour of the $\kte-\tau_0$ parameters
(decreasing of $\kte$ was not followed by increasing of $\tau_0$, as expected).
It is not clear whether this was due to the lack of data below 3 keV, which is necessary to constrain the seed photon temperature or due to  accumulation time.
Thus we  prefer to skip these measurements. 
It is also worth mentioning that \cite{oosterbroek01} performed a \sax\ analysis of GX 3+1 and Ser X--1, both sources characterized by typical \emph{soft} state spectra, fitting them  with a DBB+\comptt\ model, but they did not specify which geometry
(sphere or slab) it was assumed. We found $\alpha_{sph}=2.44^{+0.89}_{-0.59}$ and $\alpha_{sl}=0.87^{+0.42}_{-0.25}$ for GX 3+1, $\alpha_{sph}=1.41^{+0.52}_{-0.34}$ and $\alpha_{ls}=0.45^{+0.20}_{-0.12}$ for 
Ser X--1, respectively.

Looking at Fig. \ref{alpha_data} we note that for all the analyzed sources  the spectral  index $\alpha$ 
lies in a belt around $1 \pm 0.2$, a part for the case of GX 354--0 where $\alpha \sim$ 1.6.
We give a possible  interpretation of these {\it observational} results in the Discussion.

\section{A model of Comptonization region in the neutron star case}
\label{model}

The determination of spectral index $\alpha$ obtained from Comptonization of seed photons in a bounded medium
has been faced since a long time. The emerging radiation spectrum depends on several parameters such as the geometry
 of the plasma (e.g., slab or sphere), the electron temperature and optical depth, and the space distribution of
the seed photons inside the medium. Sunyaev \& Titarchuk (1980) report the value $\alpha$ obtained from the
solution of the stationary radiative transfer equation in the non-relativistic case (Fokker-Planxk approximation)
obtained as a convolution of the time-dependent equation with the time-escape probability distribution P(u) 
for the case of source photons distributed according to the first eigenfunction of the space operator.
Later \cite{t94}, \cite{ht95} and TL95 extended the results to the sub-relativistic case considering both slab and spherical
geometry. 

In order to understand what it does happen in NS LMXBs sources, one has to consider the hydrodynamical
conditions in the region between the accretion disk and the NS surface.
We will refer to this region as the transition layer (TL), often also called boundary layer. 
Actually, the production of a strong TC bump in the persistent X-ray spectra of NS LMXBs is thought to originate in 
this TL, namely the region where matter deviates from its Keplerian angular velocity in order to match that of the slowly 
spinning NS.  The radiative and hydrodynamical configuration of the TL is mostly dictated by the Reynolds number
 $\gamma$ \citep{tlm98,to99}, which is proportional to the mass accretion rate and is eventually the inverse
of the viscosity parameter of the Shakura-Sunyaev disk. In particular, $\gamma$ \ determines the radial extension 
of the TL and in turn, from the mass accretion rate,  its optical depth.
It is worth pointing out that determination of  the vertical height-scale of the TL is in fact a very complicated problem, as it should require
a complete 3D magneto-hydrodynamical treatment of the problem. 
Using the slim disk (thus vertical-averaged) equations for determining the radial thermo-hydrodynamical structure
of the TL may be in fact an issue \citep[e.g.,][]{ps01}.

The enhanced radiation and thermal pressure because of higher electron temperature are expected to increase
the vertical height-scale of the TL with H $\approx R_{ns}$. Moreover, the solution for the angular momentum
equation \citep{tlm98,to99} for Reynolds number $\gamma > 5-10$ gives a TL radial extension $\Delta R_{TL} \la 0.5 R_{NS}$ .
With these characteristic length-scales, it seems more plausible to approximate the TL geometry to a slab whose normal is 
directed along the disk plane. 
It is worth emphasizing that Haardt \& Maraschi (1993, hereafter HM93) determined the theoretical Comptonization index
$\alpha$ considering a two-phase model for accretion disks in AGN in which a hot corona is surrounding and sandwiching 
the underlying cold accretion disk. The model can in principle be applied also to the case of solar-mass BH
 sources. HM93 assume that the corona and the disk are two slabs at significant different temperatures
and put in contact each other. The authors concentrated on the case of high temperature ($\kte \ga  $ 50 keV) and
low optical depth ($\tau < $1) for the corona, so that the diffusion approximation cannot hold, unlike what
we are considering. One of the consequences of the high-temperature treatment is that electron scattering is
anisotropic with a significant fraction of the power back-irradiating the disk. In the HM93 model, the \emph{inner} 
boundary condition of the hot corona is the disk cool surface 
(with $\ktbb < $ 5 eV) with energy-dependent albedo.
Note also that in their geometry, 100\% of the disk flux is intercepted and reprocessed by the top plasma.
In the geometry considered here on the other hand, it is possible that part of the disk emission directly escapes
the system, while a fraction of its flux is intercepted by the TL. We are actually interested here on
this portion of the intercepted disk flux (see next Section).

Because of these differences between the two models, a direct comparison of
the derived theoretical results is not straightforward. The reader can refer to the paper of HM93 for further details 
in order to better understand the differences of our and their approaches.

\subsection{Energy release  in the NS transition layer}

The energy balance in the TL is dictated by Coulomb collisions with protons (gravitational energy release), while inverse Compton 
and free-free emission are the main cooling channels 
(see a formulation of this problem in the pioneer work by Zel'dovich \& Shakura 1969).   
In fact, for the characteristics electron temperature (3 keV $\la kT_e\la$ 30 keV)  and density values 
($\la 10^{-5}$ g cm$^{-3}$) of these regions for LMXBs, Compton cooling dominates over free-free emission and   
the relation between the energy flux  per unit surface area of corona  $\qcor$,  the radiation energy  density 
 $\varepsilon(\tau)$  and electron temperature $T_e$  is given by (see also TLM98)
\begin{equation}
 \frac{Q_{cor}}{\tau_0} \approx  20.2\varepsilon(\tau)T_e(\tau),
\label{energy_balance}
\end{equation}
where  $\tau_0$ is the characteristic optical depth of
the TL.

The distribution $\varepsilon(\tau)$ is obtained as a solution of the diffusion equation
\begin{equation}
\frac{d^2\varepsilon}{d\tau^2} =-\frac{3\qtot}{c\tau_0},
\label{diffusion_equation}
\end{equation}
where now $\qtot=\qcor + \qd$ is the sum of the corona (TL) and intercepted disk fluxes, respectively. The two boundary conditions for equation (\ref{diffusion_equation})
are written as
\begin{equation}
 \frac{d\varepsilon}{d\tau} \vert_{\tau=\tau_0}=0, 
\label{bc1}
\end{equation}

\begin{equation}
\frac{d\varepsilon}{d\tau}- \frac{3}{2}\varepsilon\vert_{\tau=0}=0
\label{bc2}
\end{equation}
which represent the case of albedo A=1 at the NS source ($\tau=\tau_0$) and  no diffusion emission falling from outside onto outer corona boundary  
($\tau=0$). The condition for A=1 arises from the well-established observational result of NS temperature $\ktbb \sim$ 1
keV, which implies the presence of a ionized NS atmosphere.  This is different from the case considered by HM93, where the
cool disk temperature ($<$ 5 eV) gives rise to an energy-dependent albedo with photoelectric absorption for impinging
photons with energy $ \la$ 10 keV. Another important consideration to keep in mind is that the diffusion equation 
(\ref{diffusion_equation}) is to be considered frequency-integrated. This means that we are not dealing with the
specific (energy-dependent) shape of the reflected spectrum from the NS surface, we are only dealing with 
the total energy density.

The solution for $\varepsilon(\tau)$ is then given by
\begin{equation}
 \varepsilon(\tau)=\frac{2\qtot}{c} \left[1+ \frac{3}{2}\tau_0\left(\frac{\tau}{\tau_0} -
  \frac{\tau^2}{2\tau_0^2}\right)\right].
\label{ene_vs_tau}
\end{equation}

It is worth pointing out that $d\varepsilon/d\tau > 0$  for $\tau<\tau_0$, and as 
$F_{rad} \propto d\varepsilon/d\tau$ for NS sources the radiative
force always plays against gravity, unlike the case of BH sources.

Note that  the spectra of NS sources 
both in the soft and hard state can be adequately fitted by single-temperature Comptonization models (Pazis et al. 2006, 
Falanga et al. 2005, Farinelli et al. 2008, Cocchi et al. 2010).  This observational fact  demonstrates 
that an assumption of isothermal plasma in the TL can be applicable to  X-ray data analysis 
from NS binaries.  The question is how one can estimate this average temperature of the TL which is, in fact, established by photon scattering and cooling processes. 

In order to establish this average plasma temperature $T_e$  one should estimate the mean energy density in 
the TL as 
\begin{equation}
<\varepsilon(\tau)>=\frac{1}{\tau_0}\int^{\tau_0}_0 \varepsilon(\tau) d\tau=\frac{
\qtot}{c}(2+\tau_0).
\label{average_ene}
\end{equation}

It is worth emphasizing the  similarity between equations (\ref{energy_balance}) and (\ref{average_ene}) in
our paper and equation (13) in \cite{bk80} who studied  the radiation emission due to  gas accretion onto a NS.
If we now substitute the result of equation (\ref{average_ene}) into equation (\ref{energy_balance}), after a bit of 
straightforward algebra we obtain
\begin{equation}
\frac{\kte \tau_0 (2+\tau_0)}{m_e c^2}=\frac{0.25}{1+\qd/\qcor}.
\label{ktetau}
\end{equation}

Keeping in mind the definition of the Compton parameter $Y \approx A N_{sc}$ (Rybicki \& Ligthman 1989),
where $A\sim 4kT_e/m_ec^2$ and $N_{sc}\sim$ Max$(\tau_0^2, \tau_0)$ are the average photon energy gain per scattering and average 
number of scatterings, respectively, we can
rewrite  equation (\ref{ktetau}) as follows

\begin{equation}
Y \sim \frac{1}{1+\qd/\qcor}.
\label{compton_par}
\end{equation}

Equation (\ref{compton_par}) is one of the main  points of our theoretical model and shows that in the diffusion approximation the Compton parameter, which determines
the spectral index, is just a function of  the corona and disk cooling fluxes.

\subsection{ The radiative transfer model of small variations  of index $\alpha$ during spectral state transition in NS sources 
}
As we have shown in Section 2.1, the observed spectral  index $\alpha$ of most NS LMXBs undergoes small variation around 1, 
namely  $\alpha= 1\pm 0.2$ when the electron temperature of Compton cloud varies from about 2.5 to 25 keV (see Fig. \ref{alpha_data}).  
Thus here we propose a model the spectral formation in the TL (corona) which can explain the stability of  index  $\alpha$ if  energy release in the disk is much less than that in the 
TL. Namely we show that $\alpha \approx 1+ \rm {O} (Q_{disk}/Q_{cor})$.
 
As already pointed out in classical works (Sunyaev \& Titarchuk 1980, 1985, hereafter ST85, Titarchuk 1994), spectral formation in plasma clouds of finite dimensions (bounded medium) is related to the distribution law of the number of scatterings that seed photons experience before escaping. 
If $u_{av}$  denotes the average number of photon scatterings and the dimensionless scattering number is $u=N_e \sigma_T c t$,  then the distribution law for  $u\gg u_{av}$ is given by (see ST85)
\begin{equation}
 P(u)=A(u_{av},\tau_0) e^{-\beta u}.
\label{prob_law}
\end{equation}

For  a diffusion regime when $\tau_0 \ga 1.5$,   it results $\beta=\lambda^2_1/3$, where $\lambda_1$ is the first eigenvalue of the diffusion space operator.
As reported in ST85, the eigenvalue problem for photon diffusion in a slab
with total optical depth $2\tau_0$ is derived from solution of the differential equation 
for the zero-moment intensity 

\begin{equation}
\frac{d^2J}{d\tau^2}+\lambda^2 J=0,
\label{eigenval_eq}
\end{equation}
with absorption boundary conditions $dJ/d\tau-(3/2)J=0$ and $dJ/d\tau+(3/2)J=0$,
for $\tau=0$ and $\tau=2\tau_0$, respectively.
This leads to the trascendental equation for the eigenvalue $\lambda_n$, $n=1,2, 3 ...$ 
\begin{equation}
 \rm{tan}(\lambda_n\tau_0)=\frac{2}{3\lambda_n},
\end{equation}
which has the solution for $\tau_0 \gg 1$ and $n=1$
\begin{equation}
\lambda_1= \frac{\pi}{2(\tau_0 + 2/3)}.
\end{equation}

It is important to emphasize that the same result for $\lambda_1$ is obtained by solving equation (\ref{eigenval_eq}) for a slab with total optical depth
$\tau_0$ but with reflection condition $dJ/d\tau=0$ at $\tau=\tau_0$.
This is not surprising as such a condition is actually met at the center of a symmetric slab with total optical depth $2\tau_0$ and $0 \leq \tau \leq 2\tau_0$. Thus, the same mathematical result is obtained for two different geometrical configurations. In the first case (symmetric slab with total optical depth $2\tau_0$) it represents, e.g., an accretion disk
(ST85 treatment), in our present case we are dealing with a boundary layer
with total optical depth $\tau_0$, whose asymmetry is due to the presence of a reflector
(NS surface) at one of the two boundaries. In both cases one obtains
\begin{equation}
 \beta=\frac{\pi^2}{12(\tau_0 + 2/3)^2}.
\label{beta_as}
\end{equation}

Generalizing to the case of arbitrary optical depth $\tau_0$, the diffusion
operator $L_{\rm diff}=(1/3)d^2J/d\tau^2$ is replaced by the radiative transfer operator $L_{\tau}$ applied to $J(\tau)$ (see ST85 and TL95)
which for the disk geometry is 
\begin{equation}
 L_{\tau}J = \frac{1}{2}\int^{2\tau_0}_0 J(\tau') E_1(|\tau-\tau'|) d\tau' -J,
\end{equation}
where $E_1(z)$ is the exponential integral of the first order.
In this case, the derived value for $\beta$ is \citep[][TL95]{t94}
\begin{equation}
 \beta=\frac{\pi^2}{12(\tau_0 + 2/3)^2} (1-e^{-1.35\tau_0}) + 0.45e^{-3.7\tau_0} {\rm ln}\frac{10}{3\tau_0}.
\label{beta_general}
\end{equation}

\begin{figure}
\centering
\includegraphics[width=6cm, angle=-90]{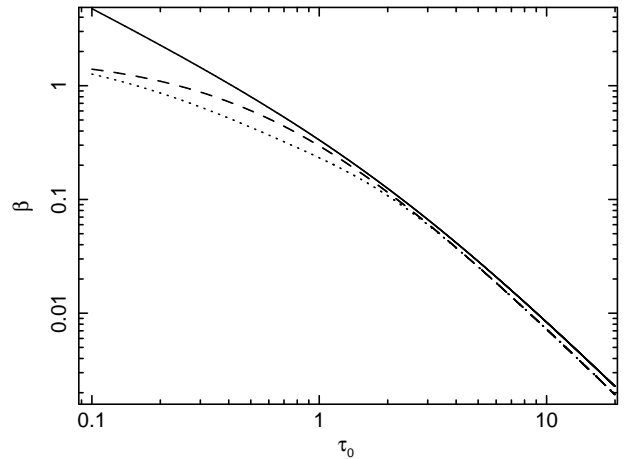}
\caption{Values of the $\beta$-parameter as a function of the optical depth $\tau_0$. The solid, dashed and dotted lines correspond to definition of $\beta$ given in equations (\ref{beta_as}), (\ref{beta_general})  and (\ref{beta_diff}), respectively.} 
\label{beta_vs_tau}
\end{figure}

Now having in mind equation (\ref{ktetau}), we introduce the parameter
\begin{equation}
\beta_{\rm diff}=\frac{1}{\tau_0 (2+\tau_0)},
\label{beta_diff}
\end{equation}
and in  Figure \ref{beta_vs_tau} we show the values of $\beta$ for the cases reported
in formulas (\ref{beta_as}), (\ref{beta_general}) and (\ref{beta_diff}) as a function
of  optical depth $\tau_0$. 
 It is possible to see that actually for $\tau_0 \ga 1.5$ all values of $\beta$ are practically close each other, 
but they deviate  for  $\tau_0 \la 1$. For example their difference is about 30\% for $\tau_0$=1.

Using  the definition of $\alpha$ (see Eq. \ref{alpha_general}), where $\beta$ is replaced by $\beta_{\rm diff}$ 
(Eq. \ref{beta_diff}), and  equation (\ref{ktetau}), we obtain  the diffusion spectral index as 
\begin{equation}
\alpha_{\rm diff}= -\frac{3}{2}+\sqrt{\frac{9}{4}+ \frac{1+\qd/\qcor}{0.25}},
\label{alpha_diff}
\end{equation}
or $ \alpha_{\rm diff}=1+0.8~ \qd/\qcor$       for $\qd/\qcor<1$.

\begin{figure}
\centering
\includegraphics[width=6cm, angle=-90]{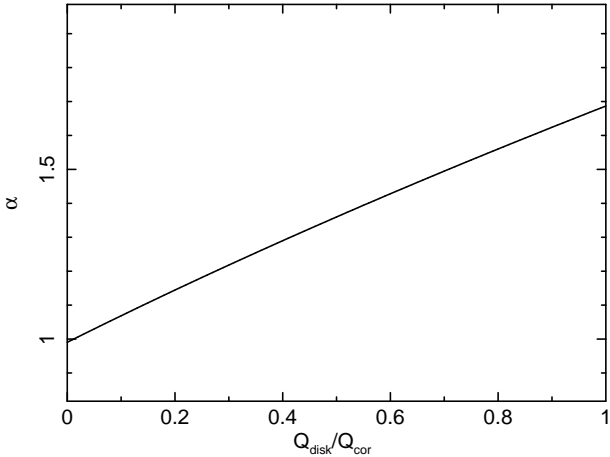}
\caption{Theoretical thermal Comptonization index $\alpha$ as a function of the ratio $\qd/\qcor$ according to equation (\ref{alpha_diff}).} 
\label{alpha_plot}
\end{figure}

Thus as it follows from Eq. (\ref{alpha_diff}), in the diffusion regime the TC spectral index can be expressed 
in terms of $\qd/\qcor$ (the intercepted disk over coronal fluxes), instead of TL electron temperature $kT_e$ and optical depth $\tau_0$
(see Eqs.  [\ref{alpha_general}] and [\ref{beta_as}]).
In  Figure \ref{alpha_plot} we present a plot of $\alpha_{\rm diff}$ as a function
of  $\qd/\qcor$, which shows that it ranges from 1 to 1.6 as $\qd/\qcor$
increases from 0 to 1. One can see the observable values of index $\alpha\sim1$ takes place if the energy 
release in the disk is much less that in TL, namely if  $Q_{disk}/Q_{cor}\ll1$.

\section{Results and discussion}
\label{results}

In this Paper we compare observational data of a sample of NS sources (Fig. \ref{alpha_data}) with the theoretical 
results  which follows from the radiative transfer model  in the diffusion approximation.
The data show  that \emph{independently of the source spectral state}, which we have parametrized through
the measured TL electron temperature $kT_e$, the spectral index $\alpha= 1\pm 0.2$.
We derived an estimate of the energy index $\alpha$ for TC spectra in NS LMXBs using an equation for the  diffusion approximation in a slab geometry with the reflection (100 \%) boundary condition, which is valid for optical depth $\tau_0 \ga 1.5$. In particular, we find that in 
this approximation it is  possible to express the value of $\alpha$ as a function  of the ratio of the flux from the accretion
disk intercepted by the corona (TL, see Eq. [\ref{alpha_diff}]) and the energy release in the corona itself.
The agreement  of the model with the data takes place when the condition $\qd/\qcor \ll 1$ holds 
(see Fig. \ref{alpha_plot}).

This behavior  of spectral index $\alpha$ actually has important consequences on  putting constraints in the accretion geometry of NS LMXBs.
In fact, as already pointed-out in the Introduction, the broad-band persistent X-ray spectra of these sources 
are usually fitted by a two-component model consisting of BB-like emission plus a strong TC bump. Both the 
cases for which the BB temperature ($\ktbb$) is lower or higher than that of the seed photons of the TC bump ($\kts$) provide
equally acceptable fits. In the first case ($\ktbb < \kts$), the origin of the direct BB component 
is attributed to the accretion disk, in the second case ($\ktbb > \kts$) to the NS surface. This dichotomy in the spectral analysis has not been overcome
for a long time. Some help in this direction has come with the discovery of the transient
X-ray tails in some of the brightest sources (see references in the Introduction): in fact, 
if the origin of the hard PL-like X-ray emission is attributed to a combined thermal plus bulk (converging flow)
effect in the innermost part of the TL close to the NS surface,
then it becomes natural to suggest that the observed direct BB photon spectrum mostly
originates in the TL/NS surface region, providing the seed photons for bulk Comptonization
\citep[see Fig. 1 in][]{farinelli07}.

The theoretical results derived in this Paper, in our opinion strengthen this
scenario. In fact, when computing the total energetic spectral budget of the sources
in the 0.1-40 keV (where most of the emission is produced) the dominating TC bump carries out more than 70\% of 
the source luminosity, the remaining part due to the direct BB component. 
If this BB originates close to the NS surface, then it is evident that the disk contribution 
to the X-ray luminosity is very small.   Given that $\qd$ in equation (\ref{alpha_diff}) represents
the flux from the accretion disk intercepted by the corona and thus is smaller or at most equal
to the directly emitted part, this eventually leads to $\alpha \sim 1$. In this framework,  
we can make some considerations about the higher value of $\alpha$ ($\sim$ 1.6) measured for GX 354--0
(see Fig. \ref{alpha_data}).
Di Salvo et al. (2000b) fitted the broad-band spectrum of the source with a BB plus \comptt\ model
(Titarchuk 1994) with $\ktbb < \kts$, a modelization corresponding to the case in which a significantly higher 
fraction of the X-ray luminosity comes from the accretion disk. This would turn of course in an enhanced
value of $\qd$, and looking at Figure \ref{alpha_plot} it is straightforward to see that increasing $\qd/\qcor$ 
corresponds to increasing $\alpha$. 
Actually, it would be interesting to see what it does happen by fitting the \sax\ spectra of GX 354--0 with  
the same model but with $\ktbb > \kts$.
 Note also in Fig. \ref{alpha_data} that for  1E 1724--3045 and two spectral states of GS 1826--238,
$\alpha \sim $1 with electron temperature $\kte \ga 20$ keV. Using equation (\ref{ktetau}) and the best-fit
values of $\kte$ reported for the two sources, with $\qd/\qcor \sim$ 0, we obtain $\tau_0 \sim$ 1.3 
for  1E 1724--3045 and $\tau_0 \sim$ 1.6-1.7 for GS 1826--238, respectively. These values of the optical
depth allow actually to deal with the diffusion approximation within degree of accuracy still satisfactory
as can be seen from Fig. \ref{beta_vs_tau}. Moreover, it is worth emphasizing that for a slab with optical
depth $\tau_0$ and inner reflection boundary condition, photons which are back-scattered from the reflecting surface
before escaping experience in fact an optical depth $\sim 2 \tau_0$, which further enhances the diffusion approximation
validity.

The other issue to point out is that the lower limit on $\alpha$ (for the extreme
case $\qd/\qcor =$0) derived by our model is 1, while there is a handful of sources for which $0.8 < \alpha < 1$.
Different reasons can in fact concur to this result. First of all, it is well known that 
multi-component modeling of X-ray spectra may have some influence in the determination of the
best-fit parameters. In particular, in the energy band where TC dominates ($\la$ 30 keV) Gaussian
emission lines around 6.4-6.7 keV are often observed and inclusion of narrow-feature component
in the model can affect the continuum parameters. Additional biases can come from the energy-band
coverage (in particular when using \xte\ data, which start from about 3 keV) and uncertainties in the calibration of the 
instrumental effective area, which may play some role in particular for spectra which are far away from being powerlaw-like.

In terms of theoretical predictions, our analytical model is intended to provide a description
of the observed stability of the spectral index but may of course have some limitations, in particular the energy-independent treatment of the radiation field and the simple slab
approximation for the TL geometry. However 
it is possible that the vertical height-scale $H$ of
the TL is higher than its radial extension, it is also likely that there is some dependence of
$H$ on the radial distance from the NS surface. Moreover, in the pure plane-parallel geometry photons are allowed to escape only through the surface of the slab, while in this case the slab  has limited extension and photons presumably can escape also from its ``walls''.

\section{Conclusions}
\label{conclusions}

 We reported results on the value of the thermal Comptonization (TC) spectral index $\alpha$ 
for a sample of LMXBs NS sources in different spectral cases and we found that a part from GX 354--0
it lies in a belt around $1 \pm 0.2$. We proposed a simple theoretical model using the diffusion
approximation where $\alpha$ is found to be a function only of the ratio of the disk and corona
fluxes. In particular when $\qd/\qcor \ll 1$ it results $\alpha \sim $1, which is  consistent
with observations. We are hopeful that our work will encourage to significantly extend the sample
of observed NS sources using archival data and observations from present and future missions,
in particular using  as broad as possible energy band, especially below 3 keV in order
to avoid biases in the spectral results.
We also claim that our model  can be helpful in solving the dichotomy related to the fact that
equally good fit are obtained for the cases in which the observed direct BB component has a temperature higher or
lower than that of the seed photons subjected to TC.

\begin{acknowledgements}
The authors are grateful to the referee whose suggestions strongly improved the quality of the paper with 
respect to the first version. This work was supported by grant from Italian PRIN-INAF 2007, ``Bulk motion Comptonization
models in X-ray Binaries: from phenomenology to physics'', PI M. Cocchi.
\end{acknowledgements}

\bibliographystyle{aa}

\end{document}